\documentstyle[12pt,aasms4]{article}
\def\spose#1{\hbox to 0pt{#1\hss}}
\def\simlt{\mathrel{\spose{\lower 3pt\hbox{$\mathchar"218$}}
     \raise 2.0pt\hbox{$\mathchar"13C$}}}
\def\simgt{\mathrel{\spose{\lower 3pt\hbox{$\mathchar"218$}}
     \raise 2.0pt\hbox{$\mathchar"13E$}}}

\def\megaw{{\bf MEGaW}}

\def\eg{{\rm e.g. }}
\def\ie{{\rm i.e. }}
\def\etal{{\rm et~al. }}

{

\slugcomment{Accepted for publication in ApJ, Part I}

\lefthead{Gibson \& Matteucci}
\righthead{On Dwarf Galaxies as the Source of Intracluster Gas}

\begin{document}

\title{On Dwarf Galaxies as the Source \\
of Intracluster Gas}

\author{B. K. Gibson\altaffilmark{1,2} and F. Matteucci\altaffilmark{2,3}}

\affil{$^1$Mount Stromlo \& Siding Spring Observatories, 
Australian National University, Weston Creek P.O., Weston, ACT, 2611, Australia}
\affil{$^2$Scuola Internazionale Superiore di Studi Avanzati, Via Beirut 2-4, 34013
Trieste, Italy}
\affil{$^3$Department of Astronomy, University of Trieste, Via G.B. Tiepolo 11, 34131
Trieste, Italy}


\begin{abstract}
Recent observational evidence for steep dwarf galaxy luminosity functions in
several rich clusters has 
led to speculation that their precursors may be the source of the majority of
gas and metals inferred from intracluster medium (ICM)
x-ray observations.  Their deposition into the ICM is
presumed to occur through early supernovae-driven winds, the resultant systems
reflecting the photometric and chemical properties of the low luminosity dwarf
spheroidals and ellipticals we observe locally.  We consider this scenario,
utilising a self-consistent model for spheroidal photo-chemical evolution
and gas ejection via galactic superwinds.  Insisting that post-wind dwarfs obey
the observed 
colour-luminosity-metallicity relations, we conclude that the bulk of the ICM
gas and metals does not originate within their precursors.  

\end{abstract}

\section{Introduction}
\label{introduction}

The existence of a hot, metal-enriched (\eg for iron, $\sim 1/2$ solar), 
gaseous component in the intracluster
medium (ICM) of clusters of galaxies has been a well-established fact for over
two decades (see the seminal review of \cite{Sar86}).  With upwards of a third
of a given cluster's total gravitational mass locked up in this gas 
(\cite{WNEF93}),
understanding the origin of this massive component has been of the utmost
importance.

That the ICM metals are the byproduct of gas that has been
processed in galaxies, and subsequently ejected, is now widely accepted --
indeed, Larson \& Dinerstein (1975)
predicted that observations of metal-enriched gas in the ICM would
be a natural consequence of Larson (1974) supernovae-driven wind model for 
elliptical galaxies.  Galactic winds are certainly the
current favoured mechanism for ejecting heavy elements 
(\eg \cite{MV88}; \cite{DFJ91}; \cite{CDPR91};
\cite{MT94}; \cite{MG95}), although it should be noted that ram pressure
stripping of the enriched interstellar media (ISMs) during the initial phase of
star formation may also play a role (\eg \cite{Sar79}, although for persuasive
arguments against the ``stripping'' scenario, the reader is directed to
\cite{W91} and \cite{RCDP93}).

Matteucci \& Vettolani (1988)
illustrated that by integrating the mass of metals ejected from
ellipticals of different initial masses over a Schechter (1976)
luminosity function
(assuming some mass-to-luminosity ratio) with faint-end
slope $\alpha=-1.25$, one could account for the mass of iron observed in
cluster ICMs as a function of cluster richness. One drawback to their early
work was that it overproduced iron in relation to
$\alpha$-elements, such
as Mg, Si, and O.  This appears to be at odds with x-ray observations of the
hot gas which
show [O/Fe]$\simgt +0.2$ and [Si/Fe]$\simgt +0.15$ (\cite{Mush94}).
Subsequent work by David \etal (1991) and Matteucci \& Gibson (1995)
alleviates this discrepancy
somewhat by allowing for a flatter than Salpeter (1955) initial mass function 
slope, by mass, of $x\approx 1$.  A more exotic mechanism for ejecting the
observed mass of metals in their appropriate abundance ratios is the
bimodal star formation scenario of Arnaud \etal (1992)
and Elbaz \etal (1995), although
a recent photometric analysis of their model Gibson (1996b)
highlights its
shortcomings, at least in its present form.

While Matteucci \& Vettolani (1988)
recovered the total ICM \it iron \rm mass, 
one of their primary conclusions was that the winds could only
contribute at most $10\rightarrow 20$\% to the total \it gas \rm mass -- the
bulk of the gas was necessarily of primordial origin.

The past few years have seen substantial advancements made, observationally, in
determining the nature of the faint end slope of cluster luminosity functions.
The gathering evidence seems to indicate that while the bright and intermediate
luminosity regime is consistent with a uniform $\alpha$ of -1.25 to -1.45 (\eg
\cite{FS91}), the faint end (\ie M$_{\rm B}\simgt -15$) dwarf spheroidal
slope (at least in rich clusters)
is significantly steeper, with values of $\alpha\approx -1.8\rightarrow
-2.2$ being favoured (\eg \cite{DPDMD94}; \cite{DPHM95}; Bernstein \etal 
1995; \cite{DPHM95}).

In a most interesting recent paper, Trentham (1994) 
speculates that the precursors
to these dwarfs which currently populate the steep faint-end luminosity
functions may in fact be the originating source for $\sim 100$\% of the
x-ray ICM gas.  Trentham (1994)
argues that if the faint-end dwarf slope is indeed
$\alpha\approx -1.9$, and the precursors eject $\sim 8\rightarrow 33$\% of
their initial \it total \rm (\ie ``initial total'' equals initial gas 
plus dark matter) mass in the form of gas (\ie $\gamma\approx 0.08\rightarrow
0.33$), after an initial burst of star
formation, then these ``dwarf precursors'' can account for most of the 
present-day ICM gas, at odds with the earlier studies which favoured a
primordial origin.

Following Trentham's (1994)
suggestion, Nath \& Chiba (1995) derived analytical expressions
for the amount of mass which can be lost by dwarf galaxies as a function
of galactic mass. They showed that the Trentham (1994)
idea was difficult to support 
if their more realistic $\gamma$ values were considered, but they also did not
adopt self-consistent models of galaxy evolution.

In light of the aforementioned observations, and subsequent work of
Trentham (1994) and Nath \& Chiba (1995),
we plan to re-examine the
potential role played by dwarf galaxies in the enrichment of gas and metals in
the ICM of galaxy clusters.  To this end, we have constructed a self-consistent,
coupled
photo-chemical evolution package suitable for the study of spheroidal star
systems in the context of a supernovae-driven wind framework.  This is the
first time, to our knowledge, that ICM enrichment models have been generated in
conjunction with the photometric evolution of the underlying galactic
population.  Details
regarding the mechanics of the population synthesis and chemical evolution
implementation are given in Gibson (1995a,b).  Earlier versions of the code were
demonstrated in Gibson (1994a,b).

We begin Section \ref{analysis} with a review of the observational constraints
for the problem at hand.  Following this, we shall introduce our favoured
``template spheroidal models'', which in turn form the basis for the subsequent
analysis of the ICM abundances and gas mass.
The discussion of our results, and summary, can be found in Sections
{\bf 3} and {\bf 4}, respectively.

\section{Analysis}
\label{analysis}

\subsection{Observational constraints}
\label{observations}

Previous models which use supernovae-driven winds from ellipticals to enrich
the ICM have had at least one major drawback -- no self-consistency check on the
implied photometric properties of resultant galaxies (\eg \cite{MV88};
\cite{PCSP89}; \cite{DFJ91}; \cite{CDPR91}; \cite{ARBVV92}; \cite{OCKF93};
\cite{MT94}; \cite{EAV95}; \cite{MG95}).  It is all very well to adjust
various input ingredients to the models in order to maximise the mass of ejecta
and/or favour specific abundance ratios, but it is imperative to test that this
has not been done at the expense of replicating the observed
colour-luminosity-metallicity relationships.

Figure \ref{fig:metallicity} shows the metallicity versus V-band luminosity
relation for dwarf through giant ellipticals.  Our model curves will be
discussed in Section \ref{template}.  The absolute magnitudes for the 
Virgo cluster ellipticals were derived assuming
$H_0=85$ km/s/Mpc, and indeed this value of the Hubble constant is assumed
throughout the remainder of the paper.  The metallicities from the
Terlevich \etal (1981) and Sil'chenko (1994)
samples were derived from Mg$_2$ line index
measurements, whereas the lower mass dwarf spheroidals (taken from \cite{S85}
compilation) are typically estimated from giant branch locations in the
spheroid's colour-magnitude diagram (\cite{S85}, and references therein).  
As an aside, the dominant elemental components of ``Z'' for these dwarfs are the
$\alpha$-elements, whereas both iron and $\alpha$-elements contribute to the
more massive systems (\cite{M94}).
Ideally one would, for
example, like to compare synthetic Mg$_2$ indices for our model giant
ellipticals against those shown in the figure, but as this is outside the scope
of our current analysis, we settle for a compromise
comparison with the global metallicity Z.  
In order to estimate the value of Z for those galaxies whose abundances were
determined with the Mg$_2$ line, (\ie all those points
which lie above [Z]=-0.6), we adopt the [Mg/Fe]=+0.25, Mg$_2$-Z calibration
of Barbuy (1994).  In light of the measurements of Mg overabundances relative to
Fe, of this order, in giant ellipticals (\cite{WFG92}), we felt that this was
justified.  Our results do not change substantially if we were to adopt the
older [Mg/Fe]=+0.00 calibration.

\placefigure{fig:metallicity}

Figure \ref{fig:colour} shows the second of our primary observational
constraints -- the colour-luminosity relation.  We will be restricting the
discussion which follows to the optical-infrared V-K colour, although it should
be understood that others, including B-V, have been considered.  The reason we
concentrate on V-K is that it provides a more valuable constraint than say B-V.
Over the range of luminosities considered in our study, B-V does not vary by
more than $\sim 0.2$ magnitudes, whereas the V-K versus $M_{\rm V}$ relation
is flatter and 
spans $\simgt 1$ magnitude.  The near vertical distribution of ellipticals in
the B-V versus $M_{\rm V}$ plane (\eg Figure 7b of \cite{AY87}) makes this
colour, by itself, a poor constraint.

The sample of giant ellipticals in Figure \ref{fig:colour} is taken from
Bower \etal (1992),
and, following their prescription, we shift the Coma data to the
Virgo scale using $\Delta({\rm V}-M_{\rm V})=3.58$ and $H_0=85$ km/s/Mpc.  
The local dwarf colours are from Thuan (1985).  Using the local dwarfs as a
constraint can be dangerous, as it is apparent that some have suffered
complicated star formation histories (\eg Phoenix, Leo I, Fornax, Carina ... 
\cite{FB94}), whereas others
(\eg Sextans, Sculptor, Draco, Ursa Minor) show little or 
no signs of star formation besides the initial
burst.  We feel reasonably safe in ensuring our model colours trace the lower
envelope of the distribution, and simply note that subsequent strong
star formation epochs will scatter the colours redward in this figure (due to
the fact that the later bursts will be taking place in pre-enriched material
from the dying stars in previous bursts, so despite being younger, their
colours will almost certainly be redder ... see, for example, Table 5A of
\cite{W94}).

\placefigure{fig:colour}

We mention in passing one final constraint pertaining to the properties of the
elliptical galaxies themselves.  As Worhey \etal (1992)
have shown, and alluded to
earlier, giant ellipticals seem to possess a magnesium to iron overabundance,
as compared to the solar ratio, with values between [Mg/Fe]$\approx
+0.2\rightarrow +0.3$, albeit with a large scatter.  
This is almost certainly due to the chemical enrichment
history of said systems being dominated by Type II SNe, as opposed to Type Ia.

Another clue to the importance of Type II SNe, not only for the role they play
in driving the stellar [Mg/Fe] to super-solar values, but also for the one they
play in the ICM itself, comes from x-ray observations of $\alpha$-element
abundances in the hot gas.  Early observational work only allowed detection of
the strongest Fe lines (\cite{MCDI76}), 
but it has become clear over the past few
years, and especially with the results from the ASCA satellite, that there is
also an $\alpha$-element overabundance with respect to iron, compared with
solar ratios, in the ICM gas, with [O/Fe]$\approx +0.2\rightarrow +0.6$ and
[Si/Fe]$\approx +0.1\rightarrow +0.5$ (\cite{Mush94})
Again, these ratios are indicative of Type II SNe dominated
origin, as opposed to SNe Type Ia.

The combination of the $\alpha$-element to iron ratios in both the ICM and the
cluster ellipticals both strongly implicate early Type II-driven winds.

As to the absolute masses of both iron and gas implied by the x-ray
observations, we refer attention to Figures \ref{fig:fig3} and \ref{fig:fig4}.
Arnaud \etal (1992)
have demonstrated that a correlation exists between the total
luminosity originating in a given cluster's elliptical+lenticular population
and the measured mass of iron residing in the hot ICM gas.
The shaded region of Figure \ref{fig:fig3} encompasses the observed scatter in
the iron-cluster luminosity relation, as reported in Arnaud (1994).  Again, a
reasonable amount of scatter exists, but the trend does appear to be real.  For
comparison, we note the location of Virgo and Abell 2199, poor and rich
clusters, respectively.  Figure \ref{fig:fig4} shows the parallel correlation,
but this time for ICM gas mass, as opposed to iron mass.  The gas masses are
claimed to be accurate to within a factor of two by Arnaud \etal (1992), 
and that
is reflected in the width of the scatter at a given luminosity.

\placefigure{fig:fig3}

\placefigure{fig:fig4}

In summary, the primary observational constraints that we must honour in a
study of this nature are: the colour-luminosity-metallicity relations
highlighted in Figures \ref{fig:metallicity} and \ref{fig:colour}, the
[Mg/Fe]$\approx +0.2\rightarrow +0.3$ overabundance in the stellar populations
of giant ellipticals, the correlation between ICM iron mass and cluster
luminosity as seen in Figure \ref{fig:fig3}, and finally, the [O/Fe]$\approx
+0.2\rightarrow +0.6$ overabundance in the ICM.
As this is the first study to couple ICM abundances with the photo-chemical
evolution of the underlying stellar population, we will demonstrate that
previous work has suffered due to their restricting of constraints to the
chemical properties alone.

\subsection{Photo-chemical evolution models}
\label{template}

We utilise the {\bf M}etallicity {\bf E}volution with {\bf Ga}lactic {\bf W}inds
chemical evolution package, \megaw (\cite{G95a}), which
is similar in spirit to that of Matteucci \& Tornamb\`e (1987), but adopts the
more aesthetically pleasing (at least in the opinion of the first 
author!) ``mass in/mass
out'' formalism, similar to that of Timmes \etal (1995), 
as opposed to the matrix
form of Talbot \& Arnett (1973).  
Diffuse dark matter halos, and their influence upon the
system's global and gaseous
gravitational binding energy, are included, following Bertin \etal (1992).  
As in Matteucci (1992),
we adopt initial dark-to-luminous masses, and radial extents, 
of 10. This means that the dark matter halos, although heavy, 
are very diffused and their effect on the potential well is 
not important, as shown by Matteucci (1992). 
Therefore, we are in a situation of almost minimal binding energy for the 
galaxies. 
As a consequence, the amounts of matter restored by galaxies into the ICM are close to the maximum ones.
Moreover, the assumed constancy of the ratio between dark and luminous matter from galaxy to galaxy is leading to a situation where
the dwarfs can contribute to a maximum amount of matter. 
In fact, there are suggestions (\cite{K90}) that the percentage 
of dark matter in dwarf 
is likely to be higher than in giant galaxies.
We do not make any assumption concerning the nature of dark matter which could be either baryonic, non-baryonic or a mixture of both. 
In any case, the nature of dark matter is not relevant to the results of this paper.

A Schmidt (1959) star formation law (\ie one in which $\psi$ varies with some
power of the gas mass) of the form
\begin{equation}
\psi(t) = \nu M_{\rm g}^k,
\label{eq:sfr}
\end{equation}
\noindent
with $k=1$, is assumed for the pre-wind (\ie $t\le t_{\rm GW}$) phase, whereas
$\psi(t)=0$ for $t>t_{\rm GW}$.  The star formation time-scale $\nu$ is
used as a free parameter in order to ensure the
colour-metallicity-luminosity relations are recovered, as will be shown in
Section {\bf 3}.  This is similar to the procedure followed by Arimoto \&
Yoshii (1987) and Yoshii \& Arimoto (1987).
Generally though, $\nu$ is found to increase
with decreasing mass in a manner reminiscent of models whose initial time-scale
for star formation is set by the mean collision time of star-forming fragments
in the proto-galaxy (\cite{AY87}).

\bf MEGaW \rm has the flexibility to use any number of input ingredient
sources.  For the purposes of this work, we have chosen a universal stellar
initial mass function (IMF) of slope $x=0.95$, consistent with the value
implicated by our earlier work (\cite{MG95}), with corresponding lower and
upper limits of 0.2 M$_\odot$ (\cite{PDR95}) and 65 M$_\odot$,
respectively.  The main sequence lifetimes come from Schaller \etal (1992),
and the remnant masses are
based upon the analytical expressions of Prantzos \etal (1993).

For the nucleosythesis
yields, we use the most recent metallicity-dependent tables of Woosley \&
Weaver (1995) for
masses $m\ge 10$ M$_\odot$ (\ie Type II SNe).  
The classic models of Renzini \& Voli (1981) are adopted for
single low and intermediate mass stars (\ie $m\le 8$ M$_\odot$).  Following
Iwamoto \etal (1994),
we assume that stars in the initial mass range $8\le m\le 10$
M$_\odot$ undergo core collapse, as opposed to thermonuclear explosion, thereby
trapping their newly synthesised metals in the resultant remnant (\ie they only
enrich the ISM via pre-collapse stellar winds).  The binary Type Ia SNe model
of Whelan \& Iben (1973)
has been included, following Greggio \& Renzini (1983).
For their yields,
we use the updated Model W7 of Thielemann \etal (1993).
We have ensured that the
present-day rate of Type Ia SNe is consistent with the observed rate in giant
ellipticals ($R_{\rm Ia}=0.07\rightarrow 0.23$ SNu, for $H_0=85$ km/s/Mpc
\cite{TCB94}), by choosing, \it a posteriori\rm, that $\sim 3$\% of the mass in
the range $3\rightarrow 16$ M$_\odot$ of the IMF gets locked into Type
Ia-progenitor binary systems (\cite{GR83}).  An average
value for $R_{\rm Ia}$ of $\sim
0.12$ SNu for the present-day rate in our giant elliptical models was found.

For gas to be expelled from a galaxy we require the thermal energy of the gas
heated by supernovae and stellar winds to exceed its gravitational binding
energy (\cite{L74}).  The stellar wind energy, while not important for giant
ellipticals, can contribute non-negligibly for low mass (\ie
$M\simlt 10^9$ M$_\odot$) systems (\cite{G94a}).  We use the energy formalism
outlined in that paper for its inclusion.  

Supernovae remnant interior thermal energy (\ie the energy assumed to be
available for driving the wind) evolution follows that of Model B$_3^\prime$ of
Gibson (1994b), which draws heavily upon the work Cioffi \etal (1988).  
Unlike the
older Cox (1972) and Chevalier (1974)
equations, which have been used exclusively
to date, Cioffi \etal (1988) 
include a sophisticated treatment of radiative cooling
in the SNR interior, as well as the effects of metallicity upon their 
evolution.  Also, virtually all previous galactic wind models have assumed that
SNRs evolve in isolation (see Model A$_0$ of \cite{G94b} -- 
in particular, that after the initial adiabatic
expansion phase a radiative cooling phase is entered and the energy 
$\varepsilon$ cools as
$\varepsilon(t)\propto t^{-0.62}$, \it ad infinitum\rm.  This is clearly not the
case, as shells either come into pressure equilibrium with the local ISM,
thereby halting the expansion cooling term, or,
more importantly, come into contact
and overlap with neighbouring expanding shells, further cooling thereafter
being negligible.  This was noted in Larson's (1974) original paper, 
but the ``evolution in
isolation'' formalism still stood (and still does in most wind models) until
the recent work of Babul \& Rees (1992) and Gibson (1994b).  
As we shall see, the
wind epochs favoured by Model B$_3^\prime$ are significantly earlier than those
predicted by Model A$_0$.  Again, further details can be found in Gibson
(1994b,1995a).

The post-$t_{\rm GW}$ energetics situation is somewhat more problematic.
Scenarios in which continuous Type Ia-driven SNe winds ensue until the
present, temporarily driven winds, or even no post-$t_{\rm GW}$ wind
whatsoever, are all feasible.  Of great significance in determining this
outcome is the amount of residual thermal energy which is left in the system
after the bulk of the gas has been ejected in the global wind at $t_{\rm GW}$.
This has been demonstrated quite graphically by Arimoto (1989) and Ferrini \&
Poggianti (1993).
We do not wish to belabour this point, and as such, we present three different
scenarios for the post-$t_{\rm GW}$ thermal evolution.  A \it minimal \rm model
in which the only gas ejected is that due to the global expulsion 
at $t_{\rm GW}$; a \it maximal \rm model in which all post-$t_{\rm GW}$ ejecta
from lower mass stars is continuously swept out of the system due to the
continued heating from Type Ia SNe; and, a \it standard \rm model, which
incorporates Arimoto's (1989) assumption that the fraction of residual thermal
energy remaining after $t_{\rm GW}$ compared with that before is $\simlt 0.01$.
These latter models usually lead to temporary winds of duration $\simlt 0.5$ Gyr
for massive systems, while for low mass systems they coincide with the \it
maximal \rm models.

In order to calculate the coupled photometric properties of our spheroids, we
have developed a simple population synthesis
package (\cite{G95b}) which, for the work
described here, used the metallicity-sensitive isochrones of Worthey (1994).
This
compilation spans $\sim$2.5 dex in [Z], and covers the primary evolutionary
stages from the zero age main sequence to the onset of the post-asymptotic
giant branch (or carbon ignition, depending upon the initial mass in question).
Luminosity weighted (V-band) metallicities $[<{\rm Z}>]_{\rm V}$, [Mg/Fe],
colours and luminosities, and final mass-to-luminosity ratios were all generated
as described in Gibson (1996b).

\subsection{Intracluster medium implications}
\label{icm}

The work of Arnaud \etal (1992) has shown that there is a direct correlation
between a cluster's elliptical population luminosity, and the measured
abundance of iron in 
the ICM.  With knowledge of a given spheroid's ejected mass
of gas (or element $i$) and its V-band luminosity, we can easily
integrate over the cluster's Schechter (1976) LF, as opposed to
working with the mass function and assuming \it a priori \rm some typical
mass-to-luminosity ratio.  By normalising to a cluster's E+S0 V-band luminosity
$L_{\rm V}^{\rm E/S0}$, we can then write the total mass of element $i$ ejected
into the ICM by galaxies of luminosity greater then $L_{\rm V}^{\rm min}$ as
\begin{eqnarray}
M_i^{\rm ej} & & = L_{\rm V}^{\rm E/S0}
\int\limits_{{L_{\rm V}^{\rm min}}/L_{\rm V}^\ast}^{\infty} 
m_i\,
\Bigl({{L_{\rm V}}\over{L_{\rm V}^\ast}}\Bigr)^\alpha 
e^{-L_{\rm V}/L_{\rm V}^\ast}\,
{\rm d}\Bigl({{L_{\rm V}}\over{L_{\rm V}^\ast}}\Bigr)\Biggr/ \nonumber \\
& & \int\limits_{{L_{\rm V}^{\rm min}}/L_{\rm V}^\ast}^{\infty} 
L_{\rm V}\,
\Bigl({{L_{\rm V}}\over{L_{\rm V}^\ast}}\Bigr)^\alpha 
e^{-L_{\rm V}/L_{\rm V}^\ast}\,
{\rm d}\Bigl({{L_{\rm V}}\over{L_{\rm V}^\ast}}\Bigr),
\label{eq:icm}
\end{eqnarray}
\noindent
where
$m_i$ is the mass of element $i$ ejected by an elliptical of
luminosity $L_{\rm V}$\footnote{
If the function $m_i$ can be written in the form of a power
law in luminosity, then equation \ref{eq:icm} can be written as a simple
incomplete gamma function (\eg \cite{MV88}; \cite{EAV95}).  This is 
sufficient \it provided \rm one restricts the fit to the high mass end of the
LF (\eg $M_{\rm g}(0)\simgt 10^9$ M$_\odot$).  
This simple power law form
does \it not \rm extend down to the dwarf (\ie $M_{\rm g}(0)\simlt
10^8$ M$_\odot$) regime.  If we were to fit a power law to the high
mass end of our standard models in Section {\bf 3}, and then blindly
extrapolate to the low mass end, we would overestimate the predicted masses of
ejected gas and metals from these dwarfs by factors of two to three.  Indeed,
the predicted ejected gas mass would actually exceed the mass of gas initially
present!
For ``normal'' $\alpha\simgt -1.45$ LFs, this unphysical
situation may not be numerically
significant, but as we are interested in exploring the relevance
of very steep ($\alpha\approx -1.9\rightarrow -2.2$) 
faint-end slopes, we feel it prudent
to integrate over the Schechter function properly.
}.  
$\alpha$ is the usual faint-end slope, with canonical values
ranging from -1.00 to -1.45
(\eg \cite{FS91}, and references therein).
$L_{\rm V}^\ast$, for the bright-end of the LF, 
is chosen to be $L_{\rm V}^\ast\equiv
1.54\times 10^{10}$ L$_\odot$ (\ie $M_{\rm V}^\ast\equiv -20.6$), again,
typical for $H_0\equiv 85$ km/s/Mpc.

The recent years have led to a revolution of sorts in our picture of the
faint-end of the cluster and field LF, and in particular, the realisation that
the faint-end is not well-described by a single slope $\alpha$.  
The ten low and medium redshift clusters in the survey of Phillips \etal (1996)
show a
surprising uniformity in that the faint-end slope brighter than M$_{\rm
V}\approx -17.3$ is consistent with $\alpha\approx -1$, whereas fainter than
this, a clear up-turn, with a power law slope 
$\alpha\approx -1.5\rightarrow -1.8$, is seen.  
The deep, very faint, LFs, from the cores of four rich clusters,
presented by De Propris \etal (1995), 
are even more extreme with a faint-end slope
$\alpha\approx -2.2$.  The models of Babul \& Ferguson (1996)
also predict a similarly
steep slope below $M_{\rm V}\approx -17.3$.
In Section {\bf 3}, we consider a number of LFs
including both the canonical single-slope faint-end, and, what appears to be,
the more appropriate two-component form,
consistent with the aforementioned observations.

One last point which should be made regarding equation
\ref{eq:icm} is the adopted lower limit on the intergals, M$_{\rm V}^{\rm min}$.
In general, or at least for all models in which the faint-end LF slope is
$\alpha\simgt -1.45$,
the results of Section {\bf 3} do not depend sensitively
upon M$_{\rm V}^{\rm min}$, and thus we typically assume 
the luminosity associated
with the lowest mass dwarf in our study (\ie $M_{\rm g}(0)=10^4$ M$_\odot$).  
As we shall see, though, for some
scenarios in which the faint-end slope of the LF is very steep,
we have to resort to more subtle
means in order to set M$_{\rm V}^{\rm min}$.  The reason for doing so is that
as Melnick \etal (1977) and Thuan \& Kormendy (1977)
have demonstrated, the \it maximum \rm
fraction of a cluster's luminosity that is tied-up in the LF, \it
below M$_{\rm V}\approx -17$\rm, is $\sim$1/4, based upon cluster
diffuse-light constraints.  Occasionally, a lower integration limit in excess of
the ``default'' must be enforced in order to ensure that we do not exceed this
$\sim$1/4 dwarf luminosity fraction.  At some level, this should not be
surprising, as one need only refer back to equation (22) of Schechter's (1976)
seminal paper to see that for $\alpha\le -2$, a regime we are
exploring in this paper, the integrated cluster luminosity diverges.


\section{Discussion}

We now present the results for our so-called ``template models'', the input
ingredients for which are described in Section \ref{template}.  Table
\ref{tbl:templatemodels} lists the output for the three post-$t_{\rm GW}$ wind
scenarios:  the extrema (continuous Type Ia SNe-driven winds to the present and
suppression of all Ia-driven winds) and the standard model (temporary winds
which die at after $\sim 0.2\rightarrow 0.5$ Gyr).  For each model we list --
column (1): the initial luminous mass in gas (in M$_\odot$) -- 
column (2): the star formation astration parameter of equation \ref{eq:sfr}
(in Gyr$^{-1}$ -- column (3): the time of the global galactic wind (in Gyrs) --
columns (4)-(6): the total mass of gas, oxygen, and iron expelled by the system
until the present epoch ($t_{\rm G}=12$ Gyr) -- columns (7)-(9): the
present-day photometric properties of interest for this paper -- column (10):
the V-band luminosity-weighted metallicity -- column (11): present-day mass in
luminous matter (star+remnants+gas) -- column (12): the total present-day
mass (luminous
plus dark, assuming initial dark to luminous mass ratio of ten) to luminosity
ratio -- column (13): the mass fraction of gas expelled by the system over its
lifetime (relative to the total initial mass) -- column (14): the final system's
gas mass fraction (relative to the final luminous mass).

\placetable{tbl:templatemodels}

Recall that the flatter $x=0.95$ IMF was chosen for consistency with our
earlier work Matteucci \& Gibson (1995),
which used predominantly the code of Matteucci \& Tornamb\`e (1987),
with the Matteucci (1992)
extensions.  Our ``minimal'' models in the table above are
similar to our earlier work, although we note that the later wind times found
in our earlier work is due mostly to our using the classic Model A$_0$
evolution for the SNe energy, as opposed to Model B$_3^\prime$ (\cite{G94b}).  
Figures \ref{fig:metallicity} and
\ref{fig:colour} illustrate that our template is successful at reproducing the
mean of the
colour-luminosity-metallicity relations observed in present-day ellipticals and
dwarfs, over $\sim$16 magnitudes in V.  As in Yoshii \& Arimoto (1987), 
the astration
parameter is treated as a free parameter in order to recover the observations.
For masses $M_{\rm g}\simgt 10^7$ M$_\odot$, $\nu\propto M_{\rm g}^{-0.1}$,
which is the expected behaviour if the initial time-scale for star formation is
set by the mean collision time of fragments in the collapsing proto-galaxy
(\cite{AY87}; \cite{M94}).

We mention in passing
that the two dwarf ellipticals in Figure \ref{fig:metallicity}
for which the model apparently overestimates the metallicity 
(NGCs 205 and 147) may themselves only be lower limits. Yoshii \& Arimoto
(1987) speculate
that this is due to that fact that
they were based upon individual red giants in the galaxies outer halos, and as
such, may not be representative of the overall metallicity of the system, and
in particular, the cores of the galaxies.

Our predicted $M_{\rm t}/L_{\rm V}$, for the standard model,
is proportional to $L_{\rm V}^{-0.07}$, 
which is flatter than the $M_{\rm t}/L_{\rm V}\propto 
L_{\rm V}^{-0.13\rightarrow -0.31}$ observational constraint from the
studies of dark matter scaling laws (\cite{K90}).  It's important to stress that
we have not attempted any ``fine-tuning'' of the input dark matter distribution
in order to recover the apparently steeper observational relation.  A global
input ratio of 10:1, both in mass and radius (dark to luminous), naturally led
to the -0.07 slope.  It is certainly easy to recover the steeper value by
systematically increasing/decreasing the dark-to-luminous initial mass ratio as
one moves to smaller/higher initial masses, but for this work we have avoided
any tinkering in that direction, especially because 
we wanted to be in the best situation for having the maximum
mass ejection from dwarf galaxies and also because
of the uncertainties relative to the amount of dark matter
in galaxies.  It is gratifying to note, 
at least, that the ratio does
increase with decreasing luminosity, contrary, for example, to what was found
in the earlier models of Yoshii \& Arimoto (1987),
which admittedly did not include any
dark matter halos (see their Figure 6).  Finally, our
$M_{\rm lum}/L_{\rm V}$ ratios for giant ellipticals are all consistent with
the observed values of $\sim 10\pm 5$ ($H_0=85$ km/s/Mpc)- 
observational values of 
the M/L ratios refer to the internal parts of ellipticals where 
the dark matter is not evident-
and increase slightly with $L_V$ ($M_{lum}/L_V \propto L_V^{0.1}$).
This behaviour is observed in real elliptical galaxies although  
the exponent is $\sim 0.2$ (\cite{BBF92}). 
A variation of the IMF from galaxy to galaxy could steepen the
predicted relation (\cite{M94}),
without any remarkable consequence on the amount of
ejected matter from galaxies. On the other hand, the trend of the 
mass to light ratio could be due to an increasing amount  and/or concentration of dark
matter with galactic mass (\cite {RC93}).

We re-iterate the point made in Section \ref{observations} regarding the
scatter in the colours for the dwarfs shown in Figure \ref{fig:colour}.
Our models trace the lower envelope of the dwarfs.  We recognise that many of
these systems show signs of intermediate age populations (\cite{FB94}), indicative
more complex star formation histories than we are capable of modelling with the
existing version of our code.  The lower envelope represents the predicted
colours for those systems which have the simple single early burst of star
formation.  Those which have subsequent phases of star formation will scatter
redward in the plot as despite being younger, they will occur in gas that has
been pre-enriched by previous generations of dying stars.

We have not tabulated the [Mg/Fe] in the underlying stellar populations, but we
state here that the giant ellipticals have a luminosity-weighted
[Mg/Fe]$\approx +0.35$, consistent with the mean
observed values from Worthey \etal (1992).
Our [Mg/Fe] values are almost constant across the luminosity range covered in
their sample, whereas their data implies a gentle increase in the value as a
function of increasing luminosity, albeit with a great deal of scatter.
This increase in the magnesium overabundance with respect to iron has been
addressed recently by Matteucci (1994).

We have also not shown the fraction of the
thermal energy at $t_{\rm GW}$ which is
due to thermalised kinetic energy from pre-SN stellar winds in massive stars
versus that due to thermalised SNe ejecta.  For giant
ellipticals this ``wind'' contribution is $\simlt 5$\%, and for low mass dwarf
spheroidals, $\sim 10\rightarrow 20$\%.
This only corresponds to an $\simlt 3$\% reduction in $t_{\rm GW}$ for giant
ellipticals ($\simlt 10$\% for dwarfs) when compared with models run with no
energy input from stellar winds.  Adjusting $\nu$ by a few percent from the
values listed in column (2) of Table \ref{tbl:templatemodels} would 
compensate easily for any minor difference in wind epochs (and hence the
resultant photo-chemical properties) for models run with and without stellar
winds, further illustrating the point made in Gibson (1994a,1996c) 
that stellar winds
do not play an important role in driving the galactic wind in 
models of this ilk.

One last input ingredient we wish to touch upon further is the initial mass
function.  Arimoto \& Yoshii (1987), David \etal (1991), and Matteucci \&
Gibson (1995),
amongst others, have
all put forth persuasive arguments for a ``flatter than Salpeter (1955)'' IMF
in elliptical galaxies.
Some arguments are based upon ICM abundances (\eg the latter two references), 
some upon the implications for the
underlying ellipticals' photometric properties (\eg the first reference).
The situation is discussed further in Gibson (1996a), but we wish to at least
draw attention to some interesting aspects of the IMF selection.

In Table \ref{tbl:imf} we show a comparison of our template $x=0.95$ IMF with
that of the steeper, canonical $x=1.35$ IMF of Salpeter (1955), each with the same
upper and lower mass limits, as before.  We only show one
initial mass ($M_{\rm g}(0)=10^{12}$ M$_\odot$), and one post-$t_{\rm GW}$
model - the ``minimal'' model - for succinctness.  The 
astration parameter $\nu$ has once again been tuned to reproduce present-day
ellipticals which follow the observations of Figures \ref{fig:metallicity} and
\ref{fig:colour}.  Similarly, the binary parameter $A$ has been chosen \it a
posteriori \rm to ensure consistency with the average present-day SNe Ia rates
mentioned in Section \ref{template} (specifically, R$_{\rm Ia}\equiv
0.12$ SNu), although the slight difference in the binary parameter $A$ (0.030
versus 0.045) is not important to the results.  
Column (1): the IMF slope, by mass, $x$ -- column (2): the astration parameter
$\nu$ -- column (3): binary mass fraction ($3\rightarrow 16$ M$_\odot$) --
column (4): the galactic wind time -- columns (5)-(7): the masses of gas, iron,
and its [O/Fe] abundance, ejected at $t_{\rm GW}$ -- column (8): the mass
fraction of gas expelled at $t_{\rm GW}$ -- column (9): [Mg/Fe] in the
underlying stellar population.

\placetable{tbl:imf}

The key thing to note from Table \ref{tbl:imf} is that selecting a steeper IMF,
but retaining the same astration parameter as for a flatter slope, leads to
colours which are significantly bluer ($\sim 0.4$ magnitudes in V-K for this
example), and metallicities which are significantly lower ($\sim 0.3$ dex),
than those observed.  A more ``gentle'' star formation scenario (the third line
in the table) is required in
order to allow the enrichment to proceed to such a level as to match the
observations.  This results in a factor of $\sim 4$ decrease in the absolute
quantity of gas mass and oxygen mass ejected at $t_{\rm GW}$, and a factor of
$\sim 2$ decrease in the iron mass ejected.  Perhaps, more importantly, the
stellar [Mg/Fe] with this ``template'' $x=1.35$ IMF is inevitably driven down
to the solar ratio, no longer in agreement with the Worthey \etal (1992) observations
of $\sim +0.2\rightarrow +0.3$.  In a related vein, the [O/Fe] ratio in the
ejected gas is also only the solar ratio.  Since this model is only marginally
greater than an $L_\ast$ galaxy, and we have not included any post-$t_{\rm GW}$
Ia-driven contribution (which could only drive this ratio further downward),
it would not bode well for any attempt at reproducing 
the oxygen-to-iron overabundance seen in the x-ray observations of the ICM gas.
It is supporting evidence such as these last points which
leads us to conclude that an IMF somewhat biased towards high mass stars (\ie a
slope of $x\approx 1$) is a necessity in the wind models
adopted to date, in agreement with our earlier study, which was based upon the
chemical properties alone (\cite{MG95}).

\placefigure{fig:fig5}

\placefigure{fig:fig6}

\placefigure{fig:fig7}

\placefigure{fig:fig8}

Using the photo-chemical properties of the ejecta, and the resultant
galaxies, shown in Table \ref{tbl:templatemodels}, we can use equation 
\ref{eq:icm} to determine the predicted ICM mass of an element (or simply 
the gas mass) originating in the elliptical/spheroidal population of a cluster
of luminosity $L_{\rm V}^{\rm E/S0}$.  This is a particularly nice aspect of
the formalism -- specifically, the predicted ejected mass is normalised to the
actual \it observed \rm cluster luminosity, and does not need to
be inferred from integrating over the mass functions, assuming some galactic
mass-to-luminosity relationship (\cite{ARBVV92}).

Figure \ref{fig:fig3} shows the results of said analysis for iron for
the \it minimal \rm
(\ie no post-$t_{\rm GW}$ contribution to the ejecta) models of Table
\ref{tbl:templatemodels}.  The solid curve shows the predicted iron ICM mass as
a function of cluster luminosity, assuming the elliptical/lenticular population
is well-described by a single Schechter (1976) luminosity function with $M_{\rm
V}^\ast=-20.6$ and $\alpha=-1.45$.  The model is only marginally consistent
with the most-metal poor observed distribution (shaded region), at a given
luminosity.  In general, the predicted iron mass is $\sim 1/2$ that of the mean
of the observed relation.

Now, let us follow the suggestion of Trentham (1994)
(and the supporting observations discussed in
Section \ref{observations}) and take the cluster dwarf spheroidal population to
be represented by a separate Schechter (1976) luminosity function from that of the
giant elliptical, with $M_{\rm V}^\ast=-16.7$ and $\alpha=-1.90$.
We now assume that 30\% of the cluster E+S0 luminosity originates in this dwarf
population, which is the upper limit set by diffuse background light
measurement in rich clusters of galaxies (\eg \cite{MWH77}; \cite{TK77}).
The remaining 70\% is associated with the giant luminosity function, with
$M_{\rm V}^\ast$ and $\alpha$ as in the previous paragraph.
As the global cluster luminosity does not change, we
are effectively shifting the available light from one part of the luminosity
function to another.

The top, heavy dotted, curve in Figure \ref{fig:fig3} shows the result of this
70/30 distribution.  The predicted ICM iron mass is now only $\sim 76$\% that
of the single luminosity function result.  The reason for this is as just
implied -- we have shifted the emphasis, somewhat, from giants, which for the
minimal model, eject $\sim 0.008$ M$_\odot$ of iron per unit luminosity, to
dwarfs, which at the low mass end, only eject $\sim 1/40$ this amount.
In this 70/30 luminosity split, the dwarfs only contribute $\sim 10$\% the
absolute mass of iron that the giants do.

Figure \ref{fig:fig4} shows the corresponding prediction for gas, in the
minimal model.  Here we see that the single luminosity function can only
account for $\sim 18$\% of the observed ICM gas at a given cluster luminosity.
The double LF is able to raise this to $\sim 23$\%, for much the same reason
that it led to lower iron -- specifically, the ejected gas mass per unit
luminosity is $\sim 5.5$ times higher at the low mass end of the minimal models
in Table \ref{tbl:templatemodels}, as compared to the giant end.  By shifting
the luminosity distribution to favour the low mass end of the spectrum, we 
do end up boosting the predicted cluster ICM gas mass which originates, but not
nearly to the level necessary to explain 100\% of the gas.

Because there is no post-$t_{\rm GW}$ contribution to the winds, and the wind
epochs are quite early ($\simlt 0.1$ Gyr), it is not surprising to note that the
predicted ICM gas [O/Fe] is $\sim +0.4$, consistent with a prominent Type II
origin to the elements (\cite{WW95}), and easily consistent with the observed
oxygen overabundance relative to iron observed by ASCA (\cite{Mush94}).
Besides the slight underproduction of iron just alluded to,
a more notable problem with the minimal model is the predicted final
fraction of gas in the system at the present-day.  From column (14) of Table
\ref{tbl:templatemodels}, we can see that the ratio of gas mass to total
luminous mass, at $t_{\rm G}\equiv 12$ Gyr, ranges from $\sim 0.6$ (dwarfs)
to $\sim 0.4$ (giants).  This is considerably higher than the observed maximum
of $\sim 10$\% (\cite{FJT85}).  Obviously then, some other mechanism has to come
into play.  We shall return to this point momentarily.

The parallel analysis for the maximal and standard models are shown in Figures 
\ref{fig:fig5} though \ref{fig:fig8}.  The maximal model overproduces iron by a
small amount ($\sim 50$\%), but is still within the observed maximum for a 
given cluster luminosity.  Because of the increased importance of the Type Ia
SNe in the post-$t_{\rm GW}$ ejecta, the [O/Fe] of the ejected gas is decreased
by $\sim 0.3$ dex, when compared with the minimal model of Figure
\ref{fig:fig3}.  The difference between the single and double luminosity
function scenarios is negligible, although the dwarfs do contribute $\sim 1/3$
of the total iron in the maximal model, as opposed to the $\sim 1/10$ we saw
in the minimal case.  

Of more interest for the maximal model, perhaps, is the result shown for the
total gas mass, in Figure \ref{fig:fig6}.  Recall that for the minimal model of
Figure \ref{fig:fig4}, the ellipticals in a cluster could contribute $\sim
18\rightarrow 23$\% of the observed ICM gas.  Now, considering the maximal
model, we see that for the single/double luminosity function scenario that the
cluster ellipticals can contribute approximately 33/38\%, of which $\sim
40$\% derives from the ``dwarfs''.
This is one of the key conclusions of our work -- \it 
even assuming an extreme scenario in which all
the gas returned by dying stars is ejected continuously to the ICM, neither the
giant ellipticals nor the dwarf spheroidals (nor their sum, for that matter)
can be responsible for all the gas observed in the ICM of galaxy clusters,
provided that we insist that the resultant galaxies reflect the photo-chemical
properties of present-day systems. \rm

Figures \ref{fig:fig7} and \ref{fig:fig8} show the results for our standard
model.  The iron mass predicted for our model ICMs lies along the median of the
observed region, but once again, the gas mass accounted for is at most $\sim
35$\%, with a similar distribution of giant/dwarf origin as was found for the
maximal model.  The predicted [O/Fe] of $\sim +0.2$ is consistent with the ASCA
observations (\cite{Mush94}).
The dwarfs are identical to the maximal models.  This is not surprising
as their shallow potential wells facilitate
the continual expulsion of gas during the
post-$t_{\rm GW}$ regime, a result reflected in the recent hydrodynamical
simulations of Wang (1995).  The giants (\ie $M_{\rm g}(0)=5\times 10^{10}$ and
$1\times 10^{12}$ M$_\odot$)
in this ``standard'' model only drove
steady winds for 0.22 and 0.43 Gyrs, respectively, beyond which the binding
energy of the returning gas from dying stars ``catches up'' with the more
rapidly cooling term for the SNe-heated gas.  At this point, one might be
tempted to re-ignite star formation, and indeed this is exactly what
Arimoto (1989) and Ferrini \& Poggianti (1993)
consider.  As this introduces
further free parameters into the picture, we decided not to pursue this option,
and we simply direct the reader to their excellent studies
of late-time star formation within the framework of the wind
model.

In this vein though, we note that the final gas fractions for our massive model
ellipticals range from $\sim 20\rightarrow 25$\%, which is still higher than
the $\simlt 10$\% expected from observations (\cite{FJT85}).  We are not overly
concerned by this apparent discrepancy, partly because of the uncertainties in
predicting the exact mass of post-$t_{\rm GW}$ ejecta, and partly because of
the neglection of post-$t_{\rm GW}$ star formation.  Indeed, the high final gas
fractions were similarly a problem with the original models of Arimoto \&
Yoshii (1987),
and as shown in Arimoto (1989), recurrent periods of star formation (at a much
reduced level to that in the initial burst) are a natural consequence of the
late-time cooling of the gas, although we re-iterate that the exact amount of
both the post-$t_{\rm GW}$ star formation and gas ejection is highly sensitive
to the assumed fraction of thermal energy in the gas immediately after $t_{\rm
GW}$.  Arimoto (1989) alleviated his $\sim 20$\% gas fraction ``problem'' without
unduly altering the present-day photo-chemical properties of the galaxies, with
limited post-$t_{\rm GW}$ star formation/gas ejection.  Again, we do not
consider such evolution in our models, but anticipate that such a scenario,
identical to his, would similarly reduce our final gas fractions without
altering our photo-chemical properties.  Still, this is an obvious avenue for
future research.

Column (13) of Table \ref{tbl:templatemodels} shows the initial mass fraction
which is ejected into the ICM in the form of gas.  Regardless of model-type,
the values range from $\gamma\approx 0.04\rightarrow 0.08$.  These compare
favourably with the lowest value considered by Trentham (1994), but are not
consistent with his higher value of 0.33.  It is important though to recall
that our values were derived from a self-consistent wind model which leads to
ellipticals (giant and dwarf) with photometric and chemical properties which
reflect those observed at the current epoch, whereas the values considered by
Trentham (1994)were somewhat arbitrary. 

\section{Summary}

In summary, we stress that by considering coupled photometric and chemical
evolution of simple spheroidal models within the framework of a
galactic wind model, the precursors to the dwarf spheroidals which we observe
today in clusters do not seem capable of providing the gas necessary to explain
the mass observed in the ICM.
Our standard model with a giant elliptical luminosity function contributing
70\% of the cluster luminosity, and the remaining 30\% originating in a low
luminosity diffuse contribution, shows that the precursor dwarfs contribute at
most $\sim 15$\% of the ICM gas, with the giants responsible for $\sim 20$\%.
The remaining $\sim 65$\% is then, by definition, ``primordial'' in origin.
This is somewhat less than the $\sim 80\rightarrow 90$\% first proposed by
Matteucci \& Vettolani (1988), so we conclude that the precursors to the dwarfs do
contribute a non-negligible amount of gas to the ICM of clusters, but that this
amount is nowhere near enough to explain \it all \rm the gas.
Our self-consistent treatment led to ejected initial
mass fractions of $\gamma\simlt
0.08$, allowing us to exclude the $\gamma=0.33$ assumption of Trentham (1994).
Whether this non-galactic wind-originating gas has resided in the cluster
since early epochs, or supplied by later gas infall, 
or simply failed to completely collapse into galactic potentials is 
still a matter under vigourous investigation 
(\eg White \etal 1993, \cite{NC95}, and references therein).

We admit that the parameter space in these wind models is large, but
we explored a much larger parameter space than presented in this work
and found that
without resorting to extraordinarily ad hoc assumptions regarding one or more
of the input ingredients, we can not envision a scenario which will eject the
required gas mass and still honour the underlying photo-chemical properties of
the resultant galaxies and the ICM abundances.
Moreover,  we stress the point that our estimate for the total 
gas ejected from galaxies is an almost 
maximal one, so that any reasonable change 
of parameters goes in the direction of further decreasing 
the amount of matter which can be restored from galaxies into the ICM, 
thus reinforcing our conclusion.

\acknowledgements

BKG acknowledges both the
receipt of a Visiting Fellowship from the Italian National
Council of Research and the subsequent warm reception received at SISSA.
The financial support of NSERC, through its Postdoctoral Fellowship program, is
gratefully acknowledged.
Helpful conversations with Neil Trentham, Roberto de Propris and
Gianni Zamorani are
likewise recognised.

\clearpage


\begin{deluxetable}{crccccrccrcrcc}
\footnotesize
\tablecaption{\label{tbl:templatemodels} Template Models\tablenotemark{a}
}
\tablewidth{0pt}
\tablehead{
\colhead{$M_{\rm g}(0)$} & \colhead{$\nu$$\;\;\;$} & \colhead{$t_{\rm GW}$} & 
\colhead{$m_{\rm g}^{\rm ej}$} &
\colhead{$m_{\rm O}^{\rm ej}$} & \colhead{$m_{\rm Fe}^{\rm ej}$} & 
\colhead{M$_{\rm V}$} & \colhead{B-V} & \colhead{V-K} & 
\colhead{$[<{\rm Z}>]_{\rm V}$} & \colhead{$M_{\rm lum}$} & 
\colhead{$M_{\rm t}/L_{\rm V}$} & \colhead{$\gamma$} & 
\colhead{$f_{\rm gas}$}
}
\startdata
\multicolumn{14}{c}{\it Minimal Model\rm} \nl
1.0e6  &  73.3 & 0.006 &  6.7e5 & 2.8e2 & 1.2e1 & -7.01  & 0.68 & 2.07 & -2.45$\;\;\;\;$ & 3.3e5  & 191$\;\;\;$ & 0.061 & 0.62 \nl
5.0e7  &  89.9 & 0.007 &  2.9e7 & 4.1e4 & 2.1e3 & -11.56 & 0.69 & 2.10 & -1.81$\;\;\;\;$ & 2.1e7  & 147$\;\;\;$ & 0.053 & 0.60 \nl
1.0e9  &  59.4 & 0.014 &  5.1e8 & 6.8e6 & 2.3e5 & -14.95 & 0.76 & 2.46 & -0.68$\;\;\;\;$ & 4.9e8  & 129$\;\;\;$ & 0.046 & 0.55 \nl
5.0e10 &  37.9 & 0.037 & 1.9e10 & 6.6e8 & 4.4e7 & -19.38 & 0.86 & 3.16 & -0.01$\;\;\;\;$ & 3.1e10 & 111$\;\;\;$ & 0.035 & 0.48 \nl
1.0e12 &  32.6 & 0.068 & 2.6e11 & 1.3e10& 9.5e8 & -22.82 & 0.92 & 3.44 & +0.30$\;\;\;\;$ & 7.4e11 &  95$\;\;\;$ & 0.024 & 0.43 \nl
\multicolumn{14}{c}{\it Maximal Model\rm} \nl
1.0e6  &  73.3 & 0.006 & 8.7e5  & 5.5e3 & 9.0e2 & -7.01  & 0.68 & 2.07 & -2.45$\;\;\;\;$ & 1.3e5  & 188$\;\;\;$ & 0.079 & 0.00 \nl
5.0e7  &  89.9 & 0.007 & 4.2e7  & 4.9e5 & 6.6e4 & -11.56 & 0.69 & 2.10 & -1.81$\;\;\;\;$ & 8.5e6  & 143$\;\;\;$ & 0.077 & 0.00 \nl
1.0e9  &  59.4 & 0.014 & 7.8e8  & 1.9e7 & 1.6e6 & -14.95 & 0.76 & 2.46 & -0.68$\;\;\;\;$ & 2.2e8  & 126$\;\;\;$ & 0.071 & 0.00 \nl
5.0e10 &  37.9 & 0.037 & 3.4e10 & 1.0e9 & 1.2e8 & -19.38 & 0.86 & 3.16 & -0.01$\;\;\;\;$ & 1.6e10 & 108$\;\;\;$ & 0.062 & 0.00 \nl
1.0e12 &  32.6 & 0.068 & 5.8e11 & 1.9e10& 2.8e9 & -22.82 & 0.92 & 3.44 & +0.30$\;\;\;\;$ & 4.2e11 &  92$\;\;\;$ & 0.053 & 0.00 \nl
\multicolumn{14}{c}{\it Standard Model\rm} \nl
1.0e6  &  73.3 & 0.006 & 8.7e5  & 5.5e3 & 9.0e2 & -7.01  & 0.68 & 2.07 & -2.45$\;\;\;\;$ & 1.3e5  & 188$\;\;\;$ & 0.079 & 0.00 \nl
5.0e7  &  89.9 & 0.007 & 4.2e7  & 4.9e5 & 6.6e4 & -11.56 & 0.69 & 2.10 & -1.81$\;\;\;\;$ & 8.5e6  & 143$\;\;\;$ & 0.077 & 0.00 \nl
1.0e9  &  59.4 & 0.014 & 7.8e8  & 1.9e7 & 1.6e6 & -14.95 & 0.76 & 2.46 & -0.68$\;\;\;\;$ & 2.2e8  & 126$\;\;\;$ & 0.071 & 0.00 \nl
5.0e10 &  37.9 & 0.037 & 2.8e10 & 7.8e8 & 8.0e7 & -19.38 & 0.86 & 3.16 & -0.01$\;\;\;\;$ & 2.2e10 & 109$\;\;\;$&  0.051 & 0.27 \nl
1.0e12 &  32.6 & 0.068 & 4.5e11 & 1.6e10& 1.8e9 & -22.82 & 0.92 & 3.44 & +0.30$\;\;\;\;$ & 5.5e11 &  93$\;\;\;$&  0.041 & 0.24 \nl
\enddata
\tablenotetext{a}{
Template Models:  \it Minimal \rm models suppress all post-$t_{\rm GW}$ SNe
Ia-driven winds; \it maximal \rm models eject all returning gas in said phase
to the ICM; \it standard \rm models include low mass systems which follow the
maximal scenario, and high mass systems which only drive winds for $\sim
0.2\rightarrow 0.5$ Gyrs subsequent to $t_{\rm GW}$.   Nucleosynthetic yield
sources are Woosley \& Weaver (1995) for high mass stars, Renzini \& Voli
(1981) for single low and intermediate mass stars, and Thielemann, Nomoto \&
Hashimoto (1993) for binary star supernovae Type Ia.  Luminosities and colours
derived from Worthey's (1994) isochrones.  Supernovae remnant thermal energy
follows model B$_3^\prime$ of Gibson (1994b), which is derived from Cioffi,
McKee \& Bertschinger (1988) and Larson (1974).  A single power law initial
mass function, by mass, of slope $x=0.95$, and lower and upper mass limits of
0.2 M$_\odot$ and 65.0 M$_\odot$, respectively, was used.  $\nu$ is the
astration parameter for the star formation rate in equation (1).  the galactic
wind time $t_{\rm GW}$ is in units of Gyrs.  A binary parameter $A=0.03$ was
adopted.  Initial dark-to-luminous mass and radial extents of 10 were chosen.
$\gamma$ is the initial mass fraction ejected in the form of gas, and $f_{\rm
gas}$ is the gaseous-to-luminous mass fraction at $t_{\rm G}=12$ Gyr.
}
\end{deluxetable}

\clearpage

\begin{deluxetable}{cccccccccccc}
\footnotesize
\tablecaption{\label{tbl:imf} $10^{12}$ M$_\odot$ Model -- IMF 
Comparison\tablenotemark{a}
}
\tablewidth{0pt}
\tablehead{
\colhead{$x$} & \colhead{$\nu$} & \colhead{$A$} & \colhead{$t_{\rm GW}$} 
& \colhead{$m_{\rm g}^{\rm ej}$} &
\colhead{$m_{\rm Fe}^{\rm ej}$} & \colhead{[O/Fe]$^{\rm ej}$} & 
\colhead{$\gamma$} & \colhead{[Mg/Fe]$_\ast$} &
\colhead{$M_{\rm V}$} & \colhead{V-K} & \colhead{$[<{\rm Z}>]_{\rm V}$}}
\startdata
0.95 & 32.6 & 0.030 & 0.068 & 2.6e11 & 9.5e8 & +0.33  & 0.024 & +0.34 & -22.82
& 3.44 & +0.30 \nl
1.35 & 32.6 & 0.045 & 0.081 & 1.2e11 & 2.7e8 & +0.26  & 0.011 & +0.22 & -23.33
& 3.01 & +0.00 \nl
1.35 &  2.9 & 0.045 & 1.450 & 7.0e10 & 5.3e8 & +0.00  & 0.006 & +0.02 & -23.32
& 3.41 & +0.30 \nl
\enddata
\tablenotetext{a}{
A comparison of the predicted ejecta composition for two different IMF slopes
for an initial gas mass of $10^{12}$ M$_\odot$.  The $x=0.95$ entry is
identical to the template model of Table 1.  A Salpeter (1955) $x=1.35$ slope
is shown for comparison -- one with the identical $\nu$ as the template, one
with $\nu$ set to recover the colours and
metallicities shown in Figures 1 and 2.  Resultant present-day SNe Ia rates are
0.11 SNu.  See text and Table 1 caption for model details.}
\end{deluxetable}

\clearpage

\figcaption[]{Observed metallicity-luminosity relation
for dwarf (open circles: Smith (1985)) and ``normal'' (open squares: Sil'chenko
1994
-- filled circles: Terlevich \etal 1981) ellipticals.  The [Mg/Fe]=+0.25, 
Mg$_2$-Z calibration
of Barbuy (1994) has been adopted for the normal/giant ellipticals.
The solid curve represents our adopted template of
models from Section \ref{template}. $H_0=85$ km/s/Mpc is assumed.
\label{fig:metallicity}}

\figcaption[]{V-K colour-luminosity relation for Virgo 
(open circles) and Coma (filled circles) cluster ellipticals and lenticulars
(Bower \etal (1992).
A Virgo distance modulus of $(V-M_V)_\circ=31.54$ is assumed, and a shift of 
$\Delta(V-M_V)_\circ=3.58$
has been applied to the Coma sample.  $H_0=85$ km/s/Mpc is assumed.
Data for local dwarfs has been taken from Thuan (1985).  The solid curve
represents our adopted template of models from Section \ref{template}.
\label{fig:colour}}

\figcaption[]{Shaded region shows the observed correlation
between the non-spiral-originating V-band cluster luminosity, and the observed
ICM iron mass, after Arnaud (1994).  
Solid curve is the single luminosity function model of slope $\alpha=-1.45$.  
Dotted lines are the components of the 
two component luminosity function model -- the lower curve is the low
luminosity dwarf spheroidal component with $\alpha=-1.90$.  The middle one is
the normal giant spheroidal population with $\alpha=-1.45$.  The heavy dotted
curve is their sum.
\label{fig:fig3}}

\figcaption[]{Shaded region shows the observed correlation
between the non-spiral-originating V-band cluster luminosity, and the observed
gas mass, after Arnaud (1994).  
Solid curve is the single luminosity function model of slope $\alpha=-1.45$.
Dotted lines are the components of the
two component luminosity function model -- the lower curve is the low
luminosity dwarf spheroidal component with $\alpha=-1.90$.  The middle one is
the normal giant spheroidal population with $\alpha=-1.45$.  The heavy dotted
curve is their sum.
\label{fig:fig4}}

\figcaption[]{Shaded region shows the observed correlation
between the non-spiral-originating V-band cluster luminosity, and the observed
iron mass, after Arnaud (1994).  
Solid curve is the single luminosity function model of slope $\alpha=-1.45$.  
Dotted lines are the components of the 
two component luminosity function model -- the lower curve is the low
luminosity dwarf spheroidal component with $\alpha=-1.90$.  The middle one is
the normal giant spheroidal population with $\alpha=-1.45$.  The heavy dotted
curve is their sum.
\label{fig:fig5}}

\figcaption[]{Shaded region shows the observed correlation
between the non-spiral-originating V-band cluster luminosity, and the observed
gas mass, after Arnaud (1994).  
Solid curve is the single luminosity function model of slope $\alpha=-1.45$.  
Dotted lines are the components of the 
two component luminosity function model -- the lower curve is the low
luminosity dwarf spheroidal component with $\alpha=-1.90$.  The middle one is
the normal giant spheroidal population with $\alpha=-1.45$.  The heavy dotted
curve is their sum.
\label{fig:fig6}}

\figcaption[]{Shaded region shows the observed correlation
between the non-spiral-originating V-band cluster luminosity, and the observed
iron mass, after Arnaud (1994).  
Solid curve is the single luminosity function model of slope $\alpha=-1.45$.  
Dotted lines are the components of the 
two component luminosity function model -- the lower curve is the low
luminosity dwarf spheroidal component with $\alpha=-1.90$.  The middle one is
the normal giant spheroidal population with $\alpha=-1.45$.  The heavy dotted
curve is their sum.
\label{fig:fig7}}

\figcaption[]{Shaded region shows the observed correlation
between the non-spiral-originating V-band cluster luminosity, and the observed
gas mass, after Arnaud (1994).  
Solid curve is the single luminosity function model of slope $\alpha=-1.45$.  
Dotted lines are the components of the 
two component luminosity function model -- the lower curve is the low
luminosity dwarf spheroidal component with $\alpha=-1.90$.  The middle one is
the normal giant spheroidal population with $\alpha=-1.45$.  The heavy dotted
curve is their sum.
\label{fig:fig8}}

\clearpage

\epsscale{1.0}
\plotone{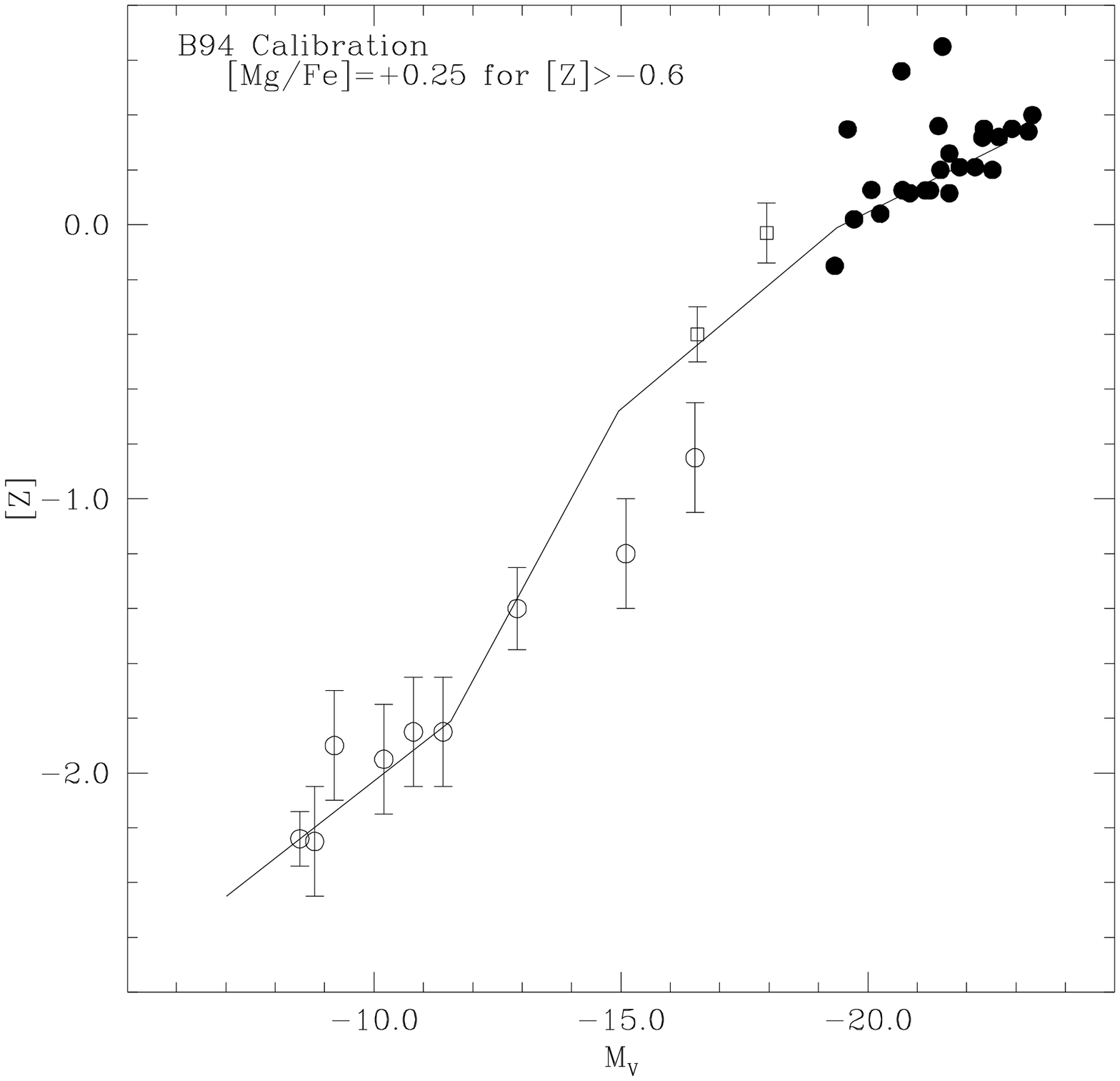}

\clearpage

\epsscale{1.0}
\plotone{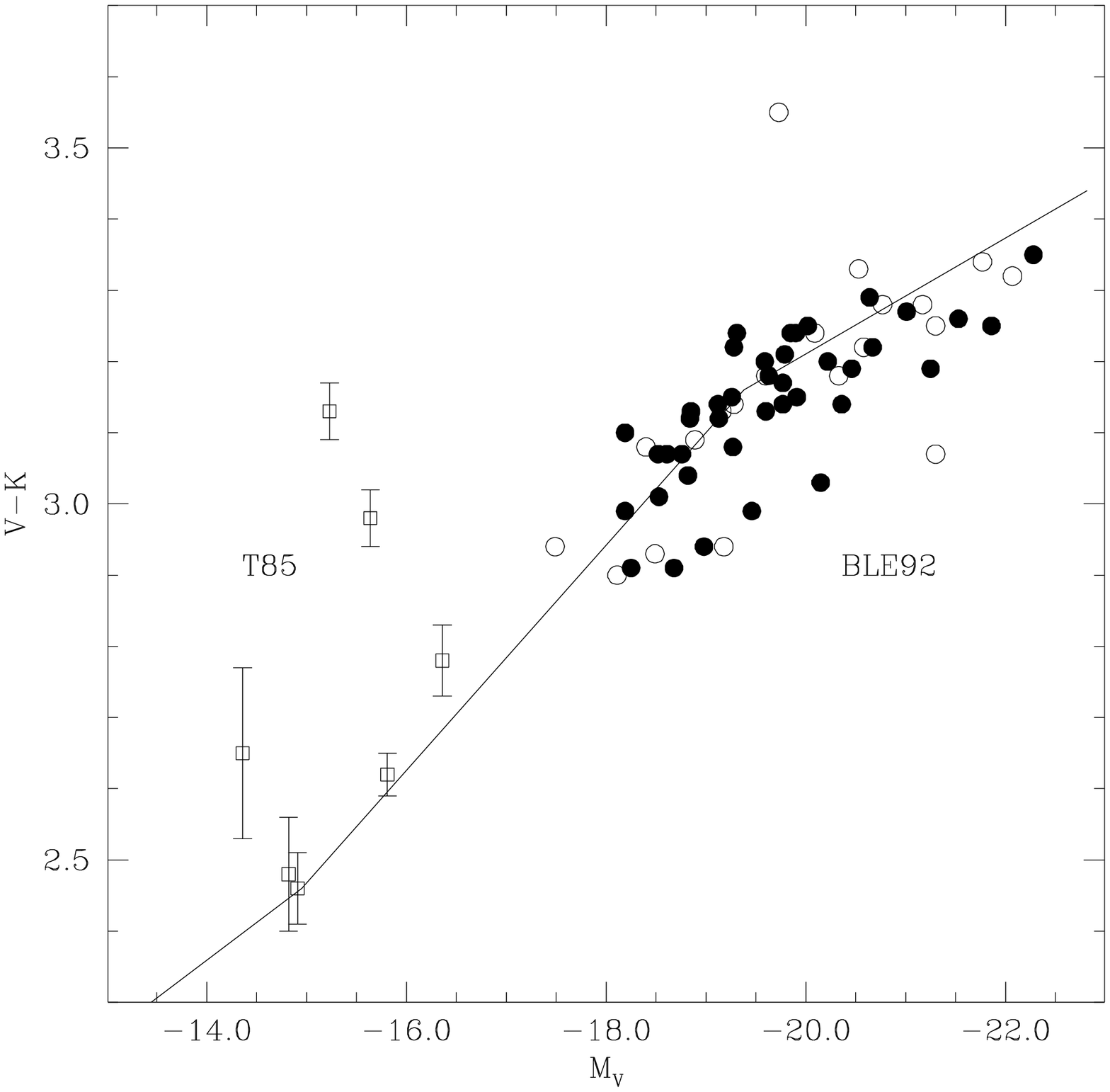}

\clearpage

\epsscale{1.0}
\plotone{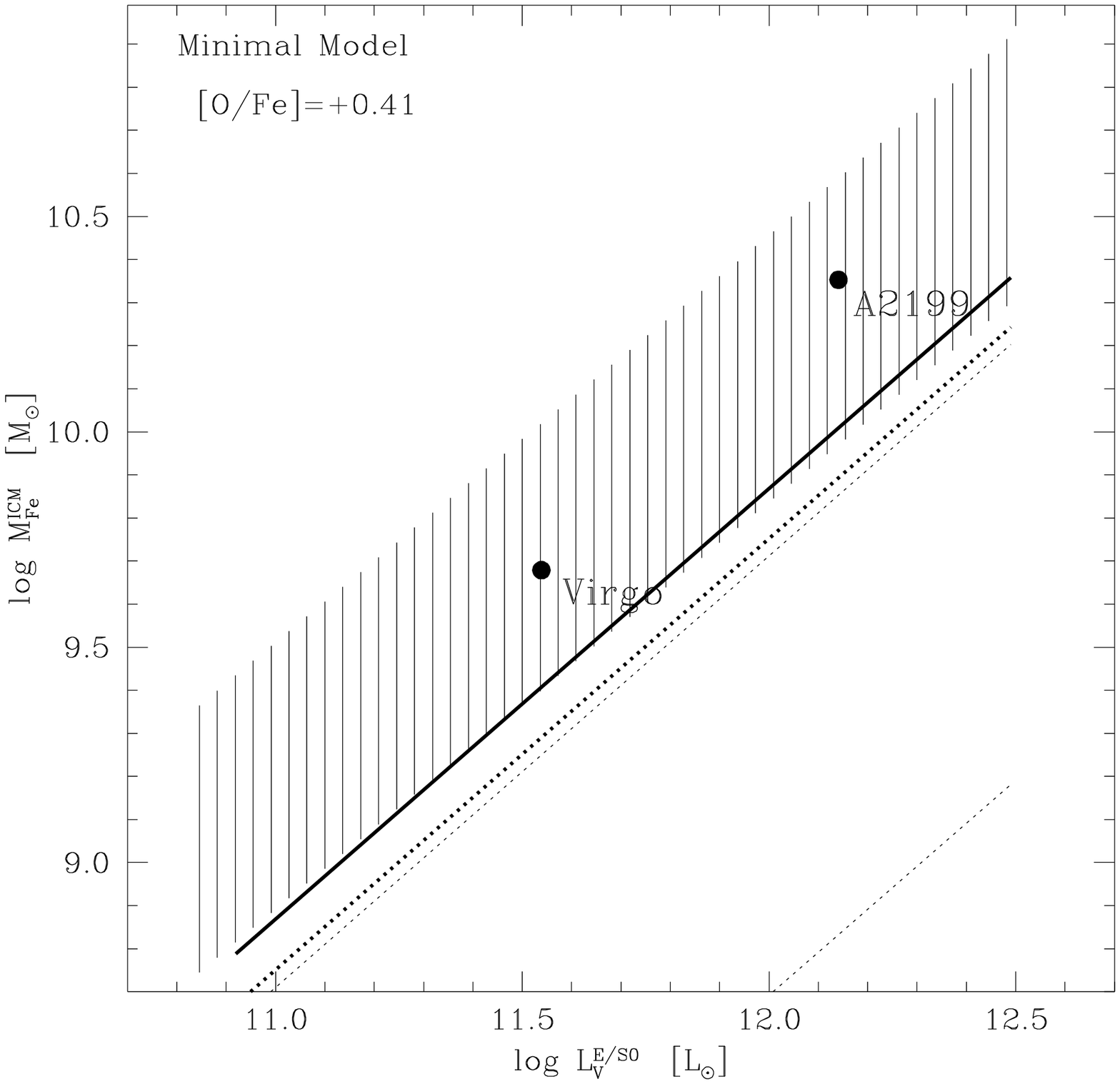}

\clearpage

\epsscale{1.0}
\plotone{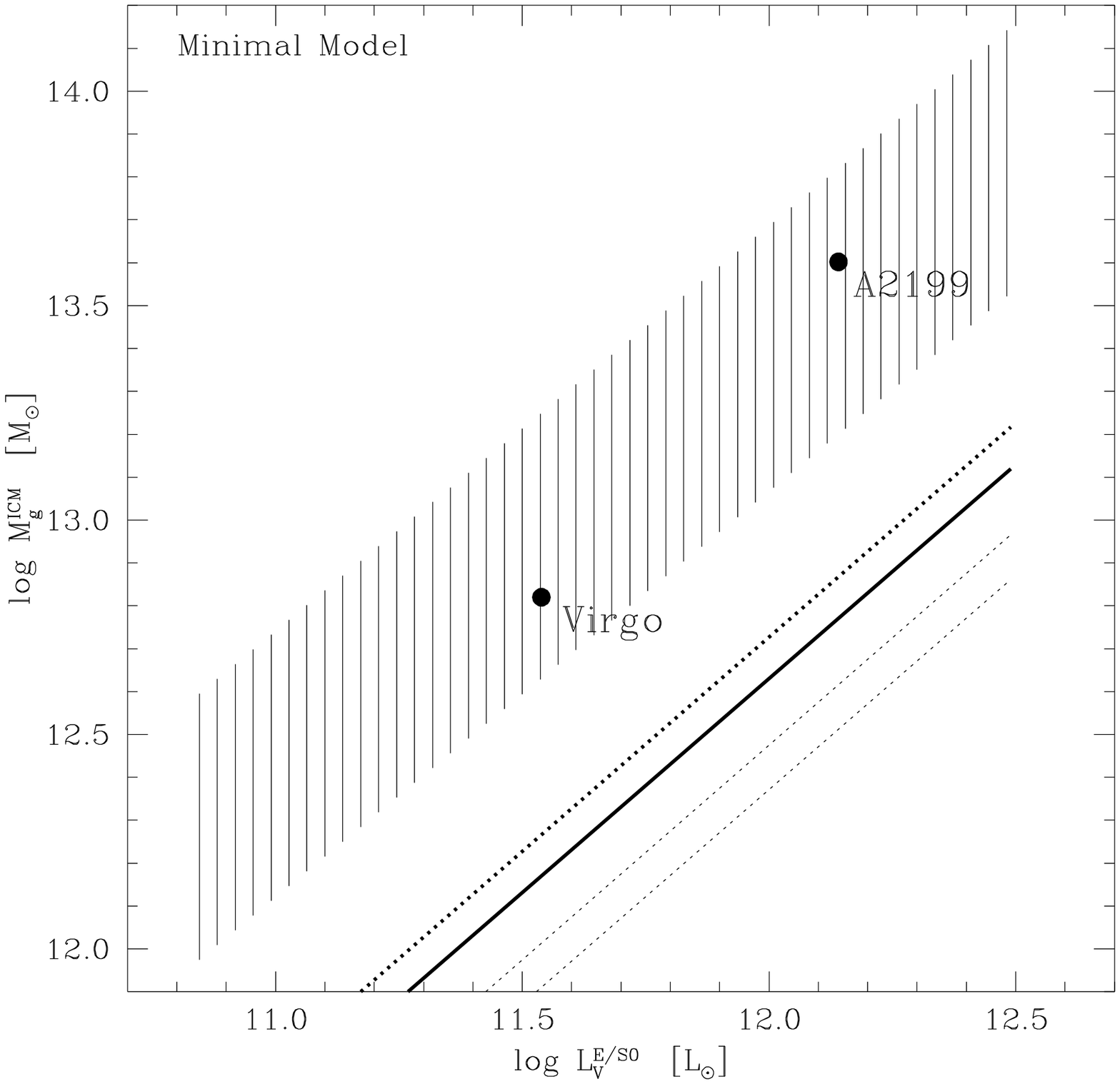}

\clearpage

\epsscale{1.0}
\plotone{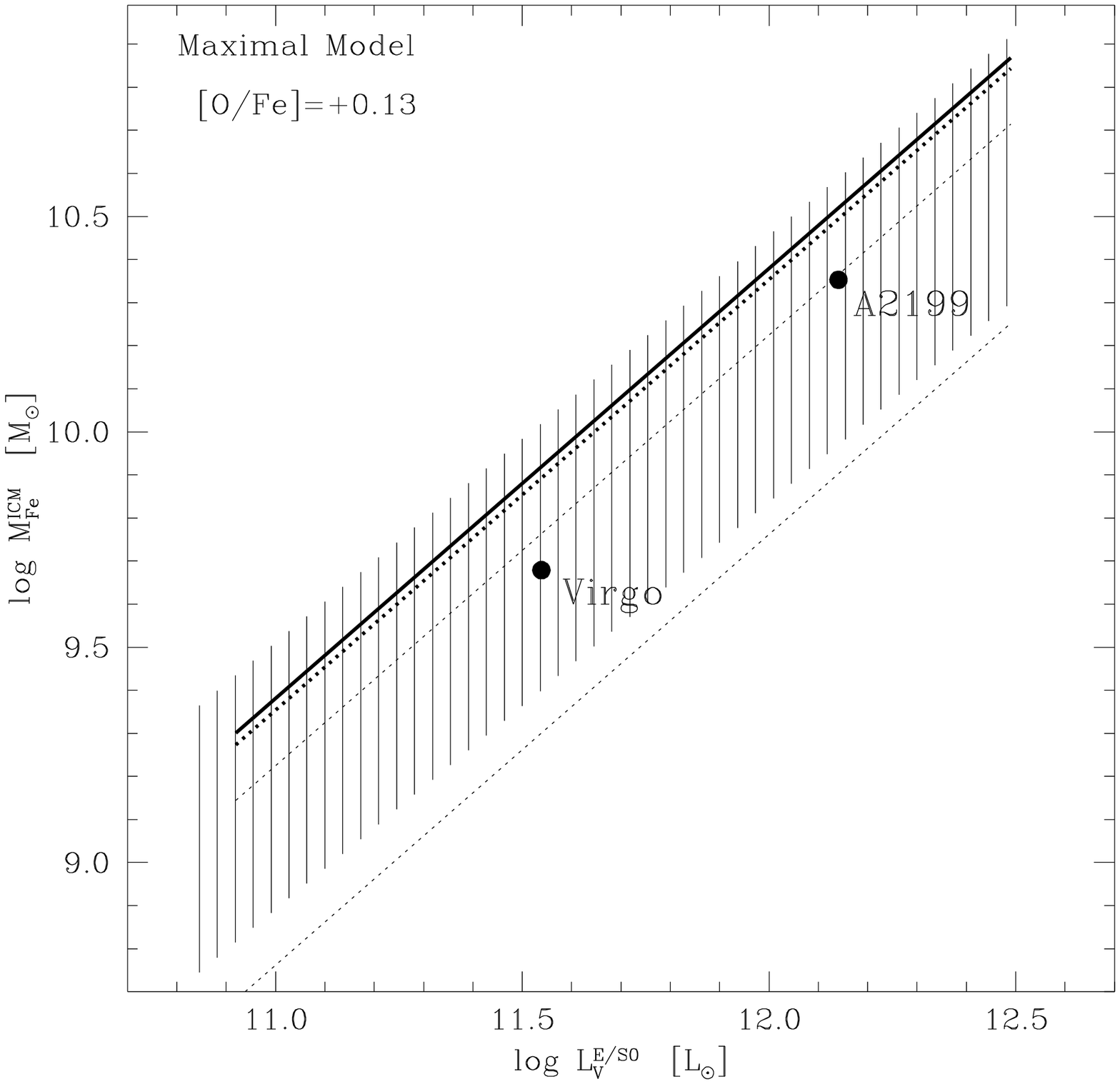}

\clearpage

\epsscale{1.0}
\plotone{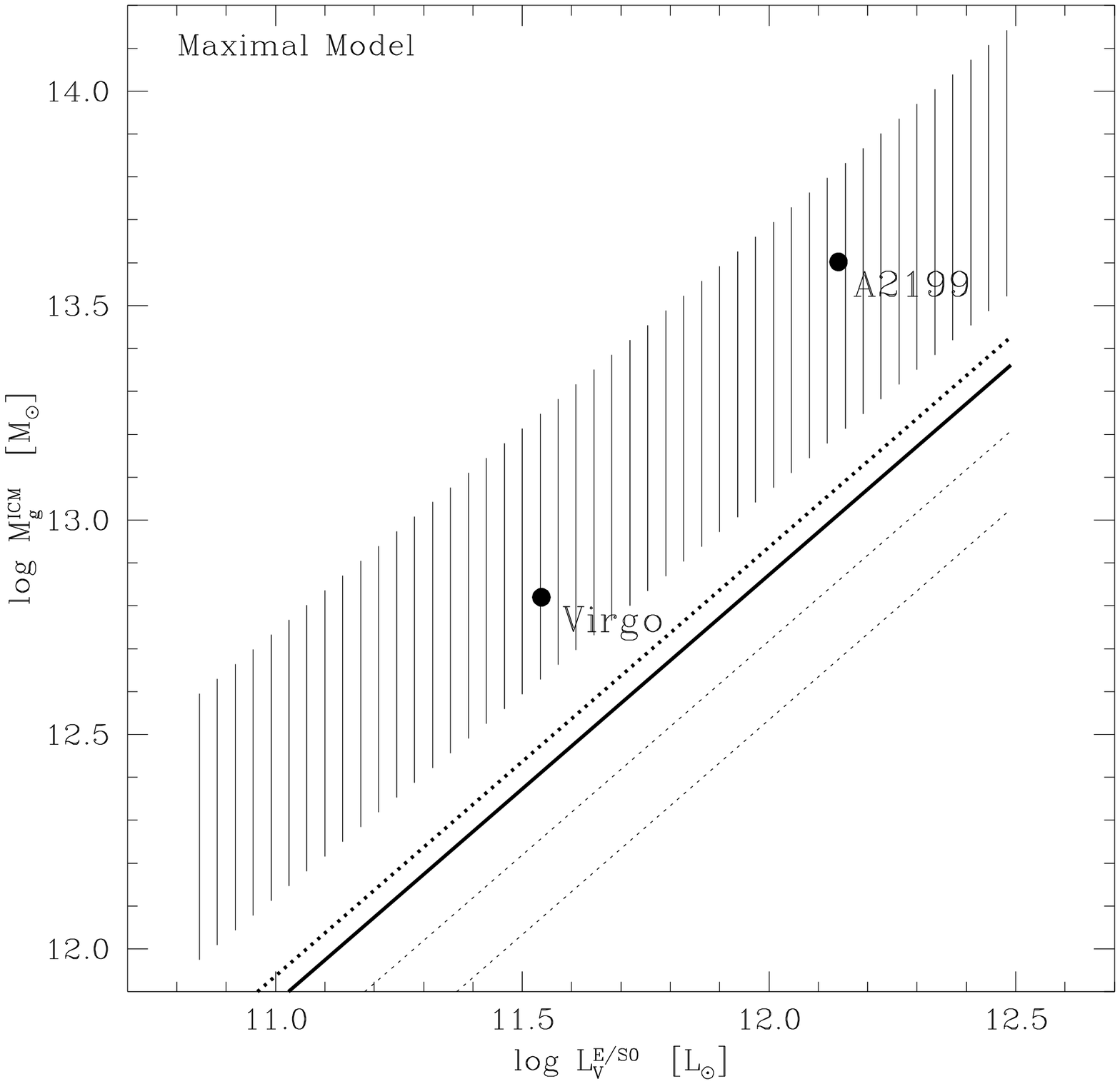}

\clearpage

\epsscale{1.0}
\plotone{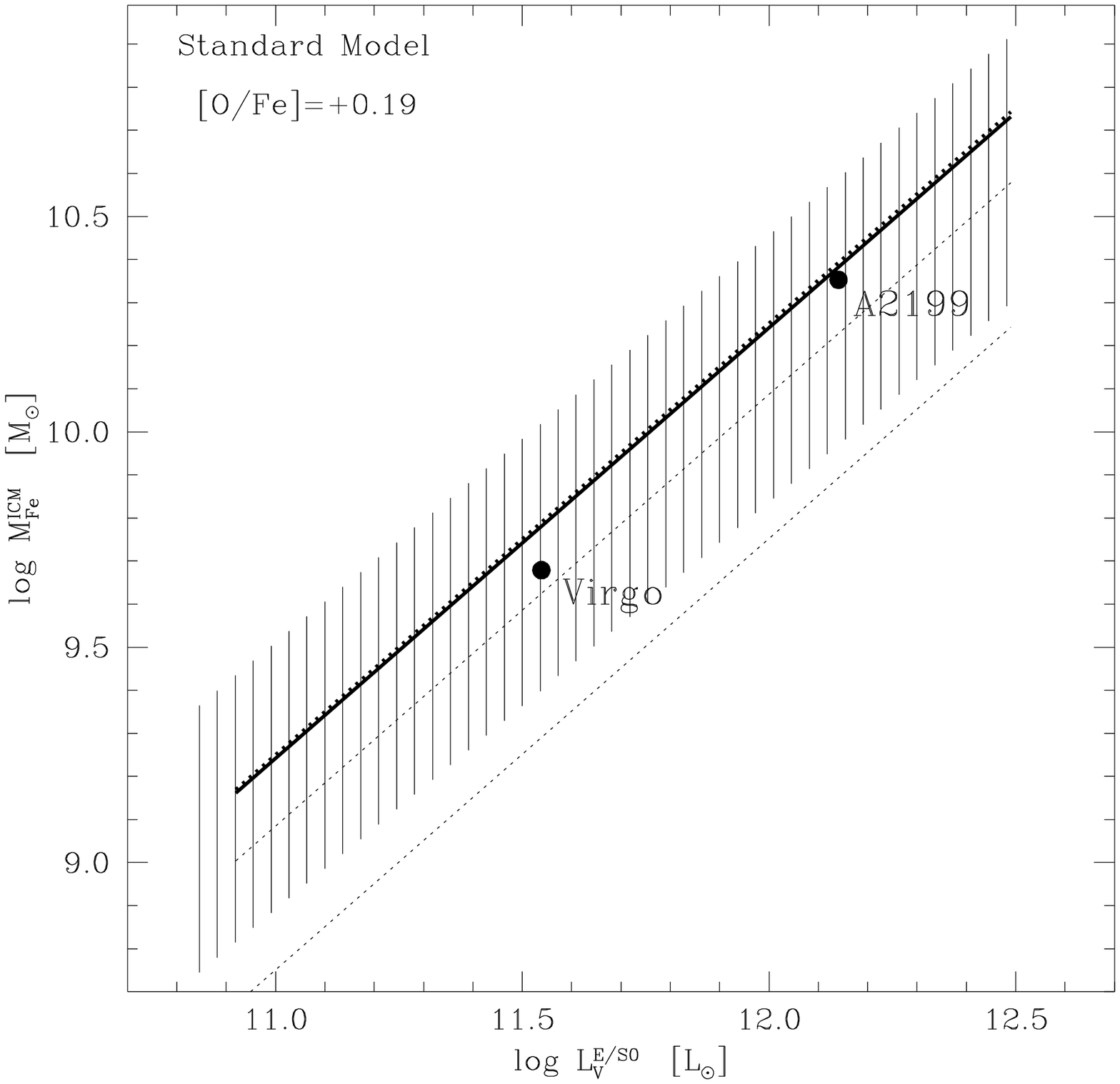}

\clearpage

\epsscale{1.0}
\plotone{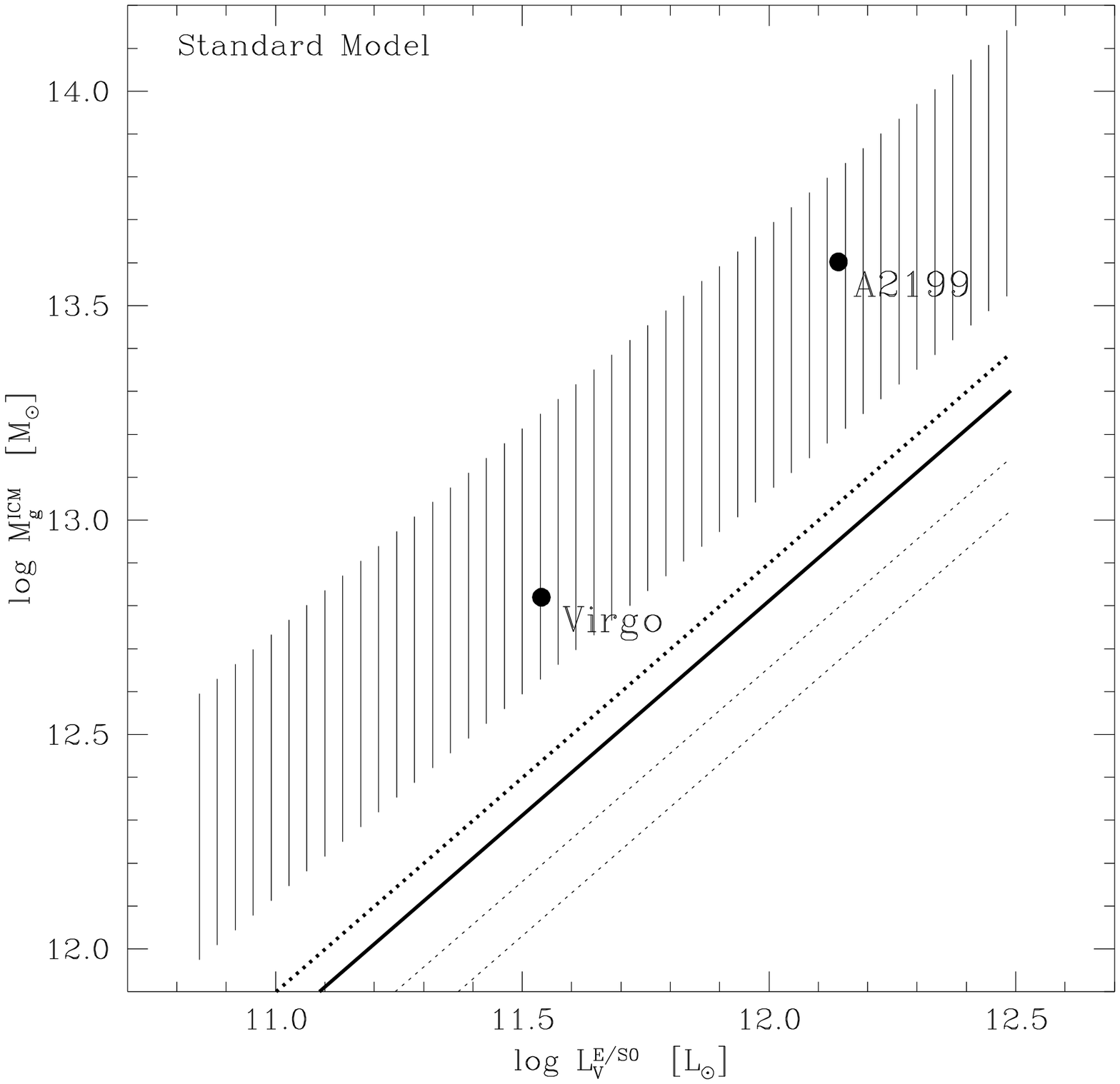}

\clearpage

\end{document}